\documentstyle[ifthen,epsfig,def]{aipproc}
\def\Journal#1#2#3#4{{#1} {\bf #2}, #3 (#4)}

\def\PLB{{\em Phys. Lett.}  B}
\def\PRL{\em Phys. Rev. Lett.}

\def\PR{{\em Phys. Rev.}}

\def\CPC{{\em Comp. Phys. Comm.}}
\def\MeV{\ifmmode {\mathrm{\ Me\kern -0.1em V}}\else
                   \textrm{Me\kern -0.1em V}\fi}
\def\GeV{\ifmmode {\mathrm{\ Ge\kern -0.1em V}}\else
                   \textrm{Ge\kern -0.1em V}\fi}
\def\ee{\mathrm{e^+ e^-}}
\def\MH{m_{\mathrm{H}}}

\def\ZZ{\mathrm{ZZ}}

\def\ee{\mathrm{e^+e^-}}

\def\WW{\mathrm{W^+W^-}}

\def\EE{\rm e^+e^-}
\def\EEZH{\rm e^+e^-\rightarrow Z\,H}
\def\zhllx{\rm ZH\rightarrow \LL X}
\def\LL{\rm \ell^+\ell^-}
\def\EEGGFF{\EE\rightarrow(\GG)\rightarrow\EE\FF}
\def\EE{\rm e^+e^-}
\def\MM{\mu^+\mu^-}
\def\TT{\tau^+\tau^-}
\def\LL{\rm \ell^+\ell^-}
\def\FF{\rm f\,\bar{f}}
\def\QQ{\rm q\overline{q}}

\def\BB{\rm b\overline{b}}
\def\GG{\gamma\gamma}
\def\ZZ{\rm ZZ}
\def\WW{\rm W^+W^-}
\def\EEWW{\EE\rightarrow\WW}
\def\EEZZ{\EE\rightarrow\ZZ}
\def\EQQG{\EE\rightarrow\QQG}
\def\QQG{\QQ(\gamma)\ }
\def\elljet{\LL+2$-$\rm jet}
\def\zhllx{\rm ZH\rightarrow \LL X}
\def\zhllbb{\rm ZH\rightarrow \LL\QQ}
\begin{document}
~~~~~~~~~~~~~~~~~~~~~~~~~~~~~~~~~~~~~~~~~~~~~~~~~~~~~~~~~~~~~~~~~~~~~~~~~~~~~~hep-ph/0302113 
%%
% The title
%
\title{Determination of the Higgs boson spin
with a linear
\mathversion{bold} $\ee$ 
\mathversion{normal}collider}

\author{M.T. Dova$^{\dagger}$, P. Garcia-Abia$^*$ and 
        W. Lohmann$^{\ddagger}$\thanks{Talk given at the 
        International Workshop on Linear Colliders,  August 26-30, 2002, 
        Jeju Island, Korea, to appear in the Proceedings.}  
        }
        \address{
        $^{\dagger}$Universidad National de La Plata, CC67, 1900 La
                   Plata, Argentina. \\
        $^*$CIEMAT, Avda. Complutense 22, E--28040 Madrid, Espa\~{n}a.\\
        $^{\ddagger}$DESY Zeuthen, Platanenallee 6, D--15738--Zeuthen, Germany}
\maketitle
%
% The abstract
%
\begin{abstract}
The  energy dependence of the production cross section
of a light Higgs boson is studied at threshold and compared to the expectations
of several spin assumptions. Cross section measurements at three
centre-of-mass energies
with an integrated luminosity of 20 fb$^{-1}$
allow the confirmation of the scalar nature of the Higgs Boson.
\end{abstract}
%
 
%
%%%%%%%%%%%%%%%%%%%%%%%%%%%%%%%%%%%%%%%%%%%%%%%%%%%%%%%%%%%%%%%%%%%%%%%%%%%%%%%
% Introduction
%%%%%%%%%%%%%%%%%%%%%%%%%%%%%%%%%%%%%%%%%%%%%%%%%%%%%%%%%%%%%%%%%%%%%%%%%%%%%%%
%
\section*{Introduction}
Spontaneous symmetry breaking in the Standard Model
leads
to one remnant
scalar particle, the Higgs boson~\cite{higgs}. 
A light Higgs boson is expected to be produced
at an $\EE$ collider of a few hundred$\GeV$
via the Higgs-strahlung process, $\EEZH$.
However,
if a particle produced in association with the Z is
detected,
the confirmation of its scalar nature will be essential to its
identification
with the Higgs boson.
Here we study the production of a Standard Model Higgs boson with a mass
$\MH = 120\GeV$ near the kinematic threshold
and compare the cross section to predictions for 
bosons
of spin 0, 1 and 2 produced in association with the Z~\cite{miller}.
%A measurement of the
%production cross section
%at three energy points
%with an integrated luminosity of 20 fb$^{-1}$ each
%Ais compared with the predictions for 
%Athe total cross section
%Aas function of the centre-of-mass energy
%A%Afor
%Abosons
%Aof spin 0, 1 and 2 produced in association with the Z~\cite{miller}.

%
%%%%%%%%%%%%%%%%%%%%%%%%%%%%%%%%%%%%%%%%%%%%%%%%%%%%%%%%%%%%%%%%%%%%%%%%%%%%%%%
% Experimental Conditions and detector simulations
%%%%%%%%%%%%%%%%%%%%%%%%%%%%%%%%%%%%%%%%%%%%%%%%%%%%%%%%%%%%%%%%%%%%%%%%%%%%%%%
%
\section*{Experimental Conditions and Detector Simulation}
The study is performed for a linear collider operated at a centre-of-mass
energies of 215, 222 and 240$\GeV$. The simulated
data statistics corresponds to an integrated luminosity
of 20 fb$^{-1}$ at each energy point.
The detector used in the simulation
follows the proposal presented in 
the TESLA Technical Design Report \cite{cdr}.
The simulation of the detector is done using
SIMDET \cite{simdet}. 

%
%%%%%%%%%%%%%%%%%%%%%%%%%%%%%%%%%%%%%%%%%%%%%%%%%%%%%%%%%%%%%%%%%%%%%%%%%%%%%%%
% Physics processes
%%%%%%%%%%%%%%%%%%%%%%%%%%%%%%%%%%%%%%%%%%%%%%%%%%%%%%%%%%%%%%%%%%%%%%%%%%%%%%%
%
\section*{Signal and background processes}

Events of the signal, $\EEZH$, are generated using PYTHIA
\cite{pythia}.
Only Z decays into electrons and muons are considered.
For the Higgs boson all decay modes are simulated
as expected in the Standard Model. 
The values of the cross section 
in the Standard Model and the expected
numbers of events corresponding to
a luminosity of 20 fb$^{-1}$ are given in Table~\ref{table:hevents}
at centre-of-mass energies of 215, 222 and 240$\GeV$.
\begin{table}[hb]
%\begin{center}
\begin{tabular}{lccc}
$\sqrt s (\GeV)$ &  $\quad$ 215 $\quad$ & $\quad$ 222 $\quad$ & $\quad$ 240  $\quad$      \\ \hline
$\sigma(\zhllx)$    & 7.2  & 12.6  & 16.8   \\
 events             & 144  &  252  &  336   \\ \hline
\end{tabular}
\caption{The cross sections and numbers
         of events expected in the Standard Model for the production 
of a Higgs boson of $\MH$ = 120$\GeV$
at centre-of-mass energies of 215, 222 and 240$\GeV$
and a luminosity of 20 fb$^{-1}$ }
\label{table:hevents}
%\end{center}
\end{table}
The following background processes are considered: $\EEGGFF$,
$\EQQG$, $\EEWW$ and $\EEZZ$. 
Initial state Bremsstrahlung is simulated by PYTHIA. Beamstrahlung
is taken into account using the CIRCE program~\cite{circe}.

%
%%%%%%%%%%%%%%%%%%%%%%%%%%%%%%%%%%%%%%%%%%%%%%%%%%%%%%%%%%%%%%%%%%%%%%%%%%%%%%%
% Methods of the measurements
%%%%%%%%%%%%%%%%%%%%%%%%%%%%%%%%%%%%%%%%%%%%%%%%%%%%%%%%%%%%%%%%%%%%%%%%%%%%%%%
%
\section*{Cross Section Measurement}

\begin{figure}[bth]
\begin{center}
\epsfig{figure=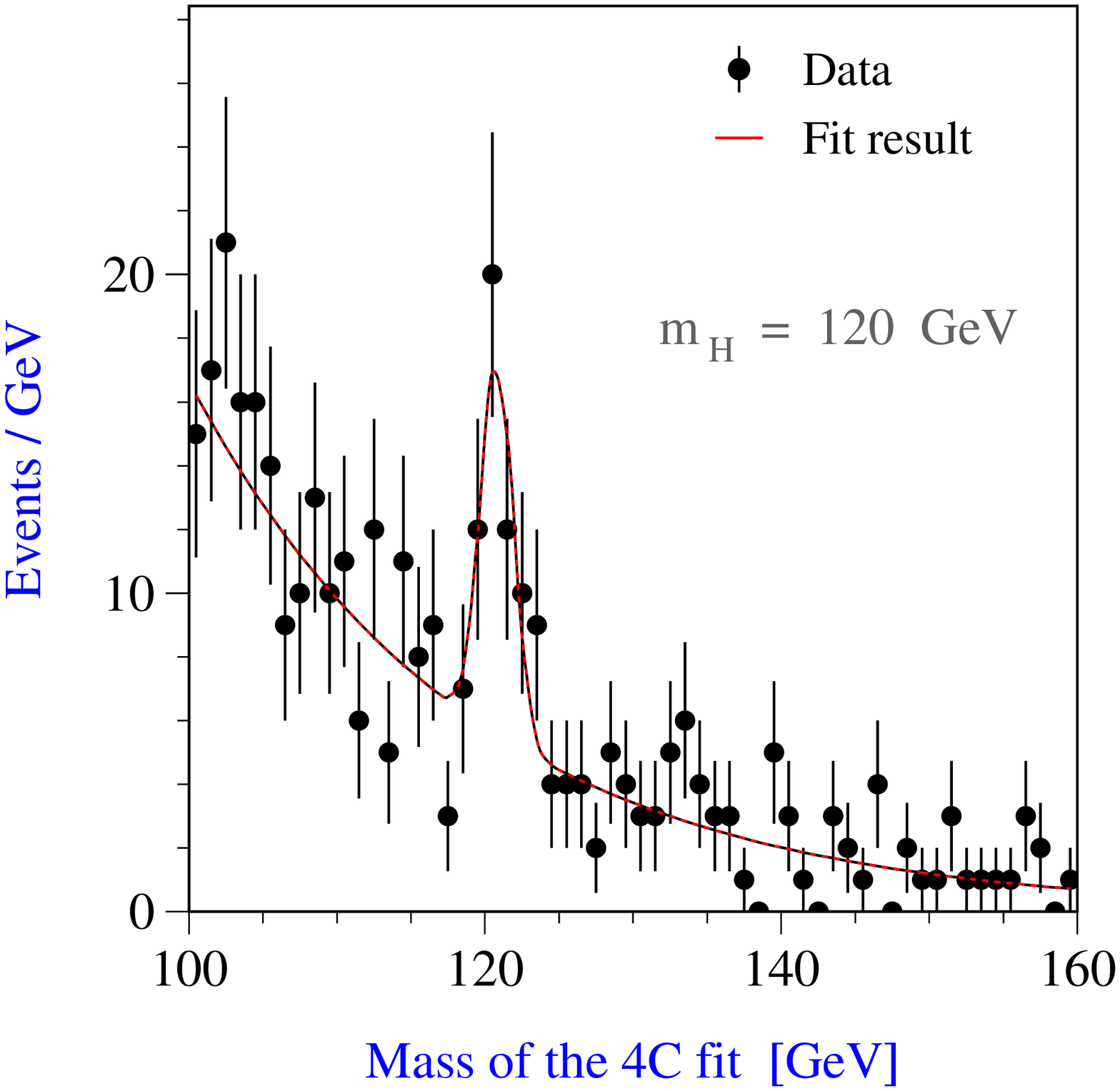,width=0.4\textwidth}
\epsfig{file=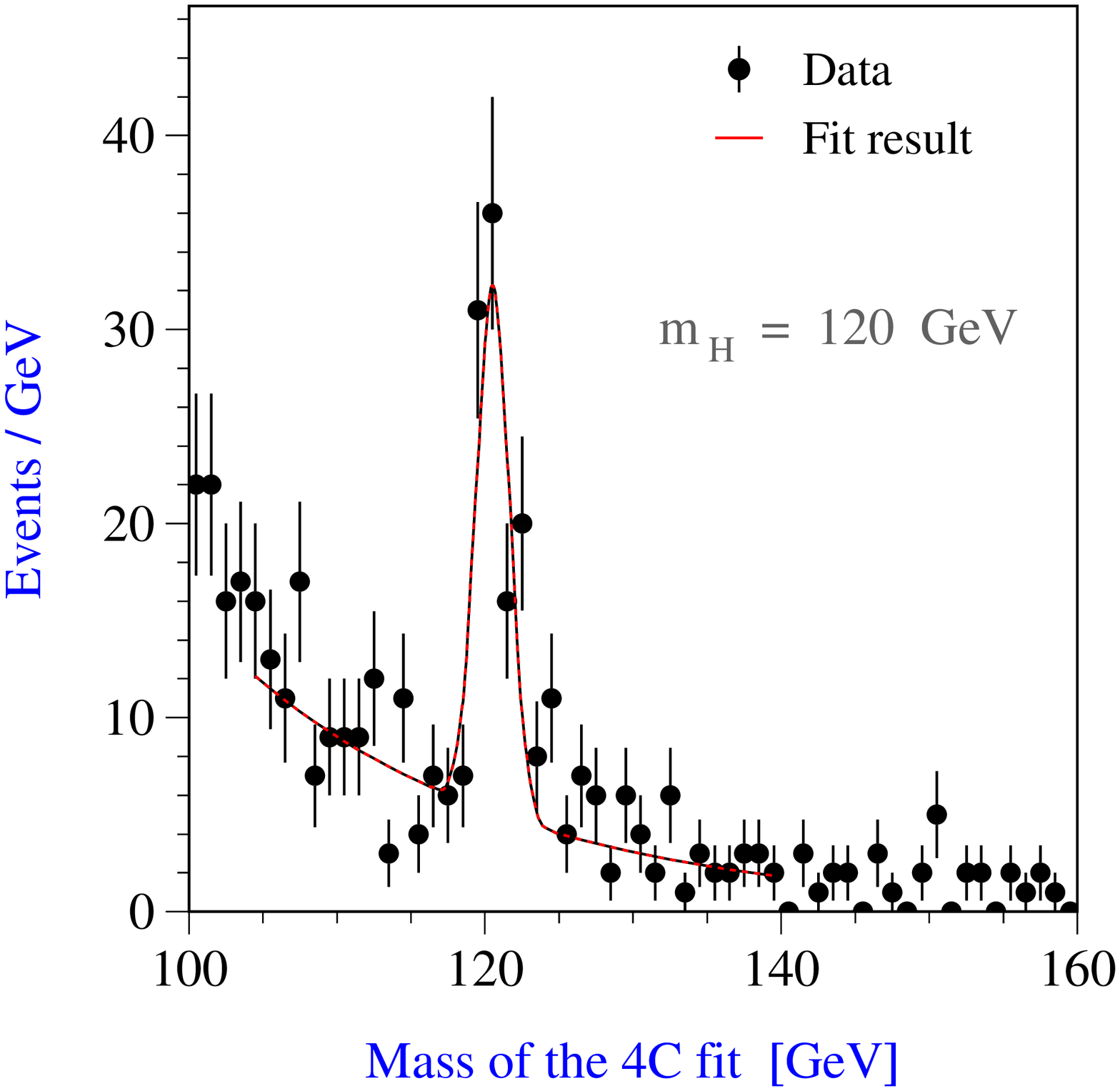,width=0.4\textwidth}
\epsfig{file=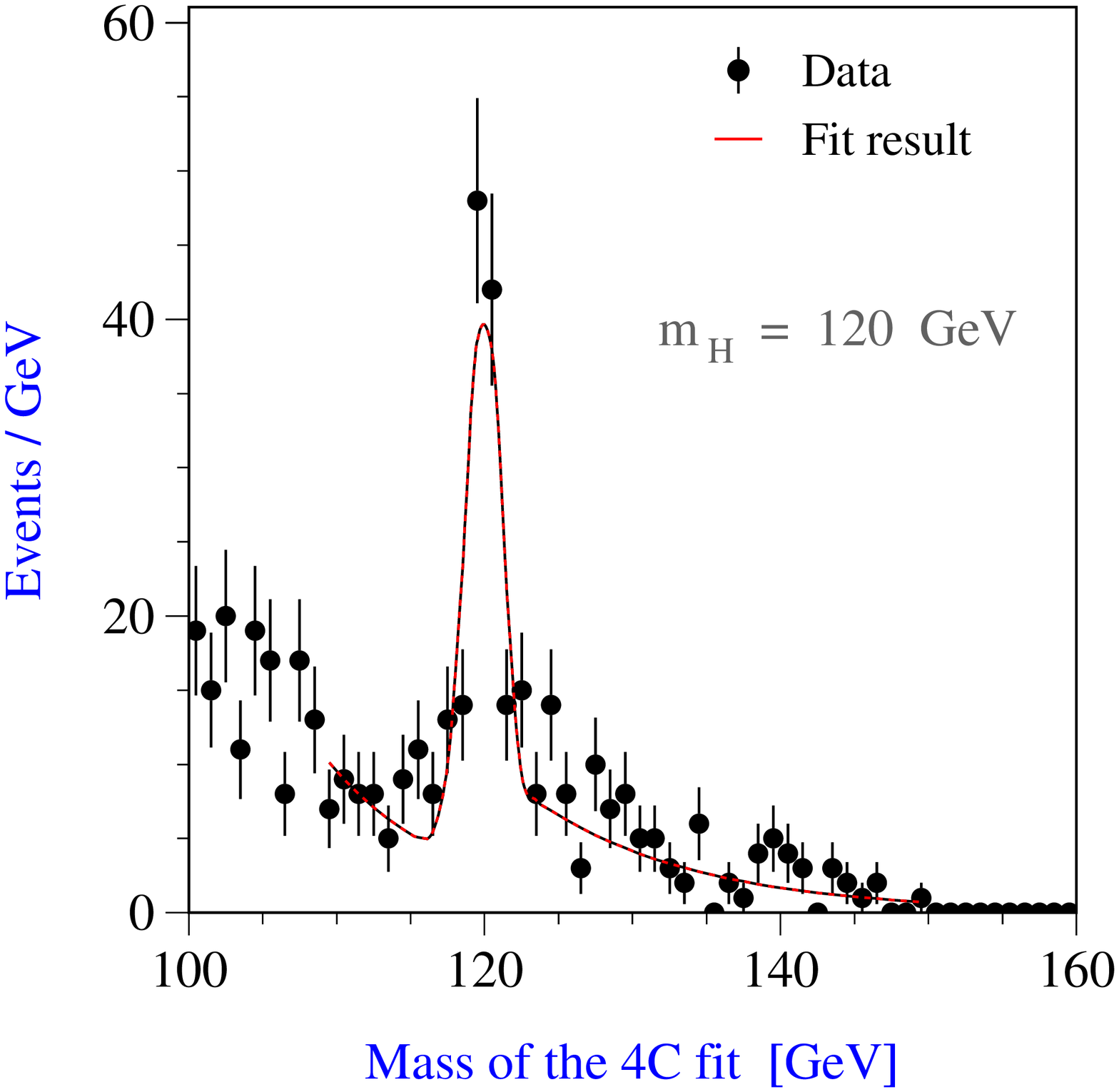,width=0.4\textwidth}
\caption{The dijet invariant mass from the $\zhllbb$ final
state after a 4C kinematic fit for $\sqrt s =$ 215
(top left), 222 (top right) and  
240$\GeV$ (bottom).}
\label{4Cmass215}
\end{center}
\end{figure}

The cross section determination is based on
the $\EEZH \rightarrow \elljet$ final state, 
where $\ell$ is an electron or muon. 
The signature 
is two isolated energetic electrons or muons and 
two jets. The identification of
electrons and muons and the formation of jets are performed
as described in Reference~\cite{glr}.
The selection efficiencies for the processes $\EEZH$ $\rightarrow \EE$
2-jet and $\EEZH$ $\rightarrow \MM$2-jet
are about 50\%.
Each event is subject of a
kinematic fit 
imposing energy and momentum conservation~\cite{blobel}.
The distributions of the two jet invariant mass 
are shown in 
Figure~\ref{4Cmass215} for centre-of-mass energies
of 215, 222 and 240$\GeV$.
The dijet mass spectra are fitted with the superposition 
of signal and background distributions.
The signal is parametrised with a gaussian
and the background is fixed to the Monte Carlo expectation.
%The uncertainties of the cross section 
%measurements, $\Delta\sigma /\sigma $,
%range between 20\% and 10\% for
%$\sqrt s =$215$\GeV$
%and $\sqrt s =$240$\GeV$, respectively.

\section*{Results}

\begin{figure}[bth]
    \epsfig{file=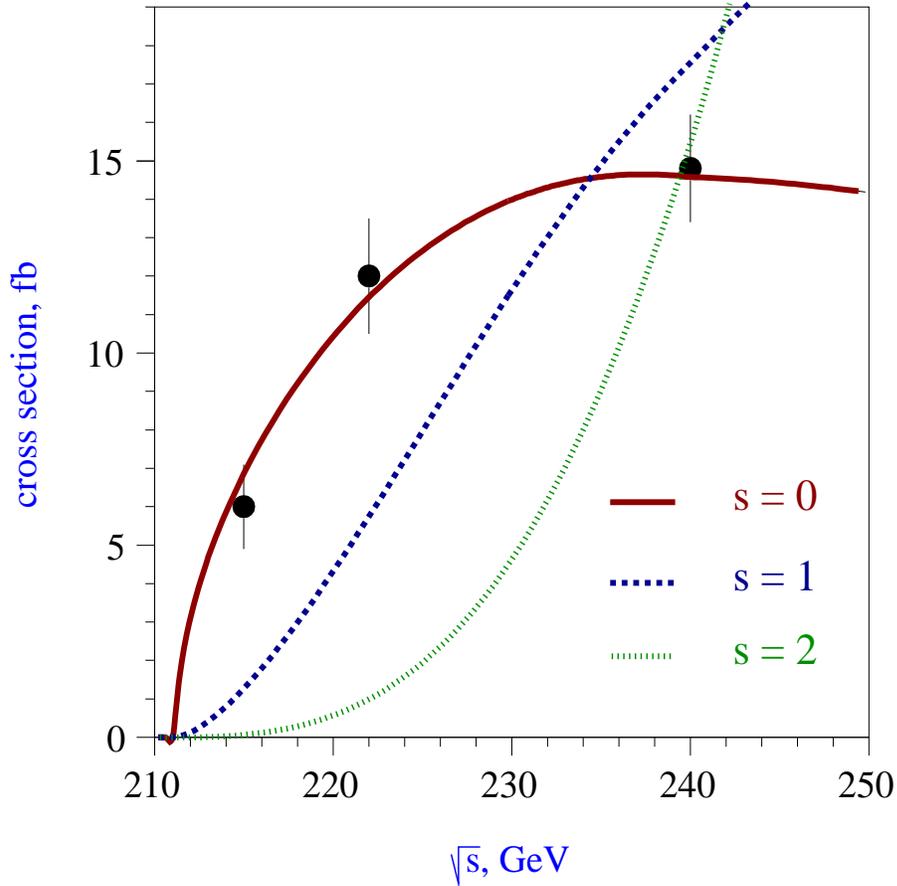,width=0.8\textwidth,
height=0.8\textwidth}
\caption{
The cross sections determined at $\sqrt s =$ 215, 222
and 240$\GeV$ (dots) and the predictions for s=0 (full line),
s=1 (dashed line) and s=2 (dotted line).
            }
\label{csvssqs}
\end{figure}

The cross sections obtained 
for the process
$\EEZH \rightarrow \elljet$ at centre-of-mass 
energies of 215, 222 and 240$\GeV$ 
are shown
in Figure~\ref{csvssqs}.
The results of fits using the predictions from models
with s=0, 1 and 2 ~\cite{miller}
with the normalisation as free parameter are also shown.
The s=0 case
is clearly distinguished from the s=1 and 2 cases.
The fit with the s=0 prediction has a good $\chi^2$ probability
and its normalisation factor equals to unity. The other fits 
have a $\chi^2$ probability of less than 10$^{-5}$.
The study of the Higgs boson production just above the kinematic threshold
shows that the measurement of the  cross section 
at three centre-of-mass energy points using a luminosity
of 20 fb$^{-1}$
allows confirmation of
the scalar nature of the Higgs boson. 
Bosons with 
other spins can be disfavoured\footnote{There are particular
scenarios
for s=1 and 2~\cite{miller}, which show a threshold behaviour similar in
shape to the s=0 one. This can be disentangled
using angular information in addition.}. 

%%%%%%%%%%%%%%%%%%%%%%%%%%%%%%%%%%%%%%%%%%%%%%%%%%%%%%%%%%%%%%%%%%%%%%%%%%%%%%%
%acknowledgement
%%%%%%%%%%%%%%%%%%%%%%%%%%%%%%%%%%%%%%%%%%%%%%%%%%%%%%%%%%%%%%%%%%%%%%%%%%%%%%%

\section*{ACKNOWLEDGMENTS}
The Authors thank
Fondation Antorchas, Argentina, for support. 

%%%%%%%%%%%%%%%%%%%%%%%%%%%%%%%%%%%%%%%%%%%%%%%%%%%%%%%%%%%%%%%%%%%%%%%%%%%%%%%
% References
%%%%%%%%%%%%%%%%%%%%%%%%%%%%%%%%%%%%%%%%%%%%%%%%%%%%%%%%%%%%%%%%%%%%%%%%%%%%%%%
%\section*{References}

\end{document}